\documentclass{elsart}
\usepackage{graphicx,natbib,amssymb}

\makeatletter
\def\elsartstyle{%
    \def\normalsize{\@setfontsize\normalsize\@xiipt{14.5}}
    \def\small{\@setfontsize\small\@xipt{13.6}}
    \let\footnotesize=\small
    \def\large{\@setfontsize\large\@xivpt{18}}
    \def\Large{\@setfontsize\Large\@xviipt{22}}
    \skip\@mpfootins = 18\p@ \@plus 2\p@
    \normalsize
}
\@ifundefined{square}{}{}
\makeatother

\begin{document}

\begin{frontmatter}
\title{Vibrational behaviour of a realistic amorphous-silicon model}

\author{J. K. Christie\corauthref{cor}\thanksref{now}}
\corauth[cor]{Corresponding author.}
\thanks[now]{Present address: The Abdus Salam International Centre for Theoretical Physics, Strada Costiera 11, 34014 Trieste, Italy}
\ead{jchristi@ictp.it},
\author{S. N. Taraskin},
\author{S. R. Elliott}
\address{Department of Chemistry, University of Cambridge, Lensfield Road, Cambridge, CB2 1EW, UK}

\begin{abstract}
The vibrational properties of a high-quality realistic model of amorphous silicon are examined.
The longitudinal and transverse dynamical structure factors are calculated, and fitted to a damped harmonic
oscillator (DHO) function.  The width $\Gamma$ of the best-fit DHO to the longitudinal 
dynamical structure factor scales approximately as $k^{2}$ 
for wavevectors $k\lesssim0.55\textrm{\AA}^{-1}$, which is above the Ioffe-Regel crossover frequency
separating the propagating and diffusing regimes, occurring at $k=0.38\pm0.03\textrm{\AA}^{-1}$.
Using the DHO function as a fitting function for the transverse dynamical structure factor (without
theoretical justification), gives a dependence of $\Gamma\propto k^{\alpha}$ with $\alpha\sim2.5$ 
for wavevectors $k\lesssim0.7\textrm{\AA}^{-1}$.  There was no evidence for $\Gamma\propto k^{4}$ 
behaviour for either polarization.
\end{abstract}

\end{frontmatter}

\section{Introduction}

The physics of the atomic vibrations present in amorphous materials has been the
subject of much investigation, see e.g.
\cite{Elliott1990,RuffleGuimbretiereCourtensVacherMonaco2006,RuoccoSette2001}.
Advances in experimental techniques such as inelastic neutron scattering 
(INS) \cite{FabianiFontanaBuchenau2005}, inelastic X-ray scattering 
(IXS) \cite{RuoccoSette2001}, hyper-Raman scattering 
\cite{SimonHehlenCourtensLongueteauVacher2006} and terahertz
time-domain absorption spectroscopy 
\cite{TaraskinSimdyankinElliottNeilsonLo2006} have allowed vibrations with 
frequencies of the order of one terahertz (1 THz $\sim4.1$ meV/$\hbar$) to be 
examined.  Vibrations at these frequencies are of particular interest, as these 
frequencies are comparable to those of the boson peak (an excess of vibrational 
modes as compared to the Debye elastic waves) \cite{TaraskinElliott2002} and of the 
Ioffe-Regel (IR) crossover between propagation and 
diffusion \cite{RuffleGuimbretiereCourtensVacherMonaco2006,TaraskinElliott2000,TaraskinElliott20002}
observed in amorphous materials.

Vibrations in amorphous materials can be characterised by the spectral density.  The
spectral-density operator is $\hat{\mathbf{A}}(\epsilon)=\delta(\epsilon-\hat{\mathbf{D}})$,
where $\epsilon=\omega^{2}$ is the squared vibrational frequency,
and $\mathbf{D}$ is the dynamical matrix \cite{EhrenreichSchwartz1976}.  
In the plane-wave basis, the matrix elements of the spectral-density operator 
$\hat{\mathbf{A}}_{\mathbf{k}\beta}(\epsilon)=\langle\mathbf{k}\beta|\hat{A}|\mathbf{k}\beta\rangle$,
where $\mathbf{k}$ is the wavevector and $\beta$ is the polarization of the plane wave,
are proportional to the dynamical structure factor $S_{\beta}(\mathbf{k},\omega)$ \cite{TaraskinElliott2002}, 
which, for the longitudinal polarization ($\beta=L$), is the quantity measured by INS or IXS \cite{RuoccoSette2001}.

This proportionality is straightforward to see.  Into the expression for $\hat{\mathbf{A}}_{\mathbf{k}\beta}(\epsilon)$ 
above, the resolution of the identity in the basis of the eigenvectors 
$\{|\mathbf{e}_{i}\rangle\}$ of the dynamical matrix $\mathbf{D}$ is substituted to give:
\begin{equation}
\hat{\mathbf{A}}_{\mathbf{k}\beta}(\epsilon)=\sum_{ij}\langle\mathbf{k}\beta|\mathbf{e}_{i}\rangle
\langle\mathbf{e}_{i}|\delta(\epsilon-\hat{\mathbf{D}})|\mathbf{e}_{j}\rangle
\langle\mathbf{e}_{j}|\mathbf{k}\beta\rangle,
\end{equation}
where $i$ and $j$ are labels for the eigenvectors, which are normalized to unity,
$\langle \mathbf{e}_{i}|\mathbf{e}_{j}\rangle = \delta_{ij}$.  This expression is equal to
\begin{equation}
\label{eq:Ake}
\hat{\mathbf{A}}_{\mathbf{k}\beta}(\epsilon)=\sum_{i}|\langle\mathbf{e}_{i}|\mathbf{k}\beta\rangle|^{2}
\delta(\epsilon-\epsilon_{i}),
\end{equation}
where $\epsilon_{i}$ is the energy of eigenmode $i$. Transforming equation \ref{eq:Ake} into frequency space,
via $\epsilon=\omega^{2}$, gives
\begin{equation}
\label{eq:Akw}
\hat{\mathbf{A}}_{\mathbf{k}\beta}(\omega)=\frac{1}{2\omega}\sum_{i}|\langle\mathbf{e}_{i}|\mathbf{k}\beta\rangle|^{2}
\delta(\omega-\omega_{i}),
\end{equation}
where $\omega_{i}$ is the vibrational frequency of eigenmode $i$. Equation \ref{eq:Akw} is valid for 
a plane wave of any polarization $\beta$.  
At high temperature, and in the single-excitation approximation, 
the dynamical structure factor for the longitudinal polarization, $\beta=L$, is defined 
as \cite{RuoccoSetteDiLeonardoMonacoSampoliScopignoViliani2000}:
\begin{equation}
\label{eq:SLkw}
S_{\beta=L}(\mathbf{k},\omega)=\frac{k_{B}Tk^{2}}{m\omega^{2}}\sum_{i}|\langle\mathbf{e}_{i}|\mathbf{k},\beta=L
\rangle|^{2}\delta(\omega-\omega_{i}),
\end{equation}
where $k_{B}$ is Boltzmann's constant, 
$T$ is the temperature, and $m$ is the atomic mass, and the Debye-Waller factor has been ignored 
(as in \cite{TaraskinElliott2002}).  This is clearly proportional to the 
spectral-density operator $\hat{\mathbf{A}}_{\mathbf{k}\beta}(\omega)$ in equation \ref{eq:Akw}, with
polarization $\beta=L$.
By defining a plane wave as 
\begin{equation}
|\mathbf{k}\beta\rangle=\frac{1}{\sqrt{N}}\sum_{j}{\bf n}_{\mathbf{k}\beta}e^{-\rm{i}\mathbf{k}\cdot\mathbf{r}_{j}},
\end{equation}
where $N$ is the number of atoms, ${\bf n}_{\mathbf{k}\beta}$ is the polarization vector of the plane wave, 
and $\mathbf{r}_{j}$ is the position vector of atom $j$, and substituting this into equation \ref{eq:SLkw},
an expression for the dynamical structure factor for any polarization $\beta$ can be derived as
\begin{equation}
\label{eq:dsf}
S_{\beta}(\mathbf{k},\omega)=\frac{k_{B}Tk^{2}}{m\omega^{2}}\sum_{i}
\vert\sum_{j}({\bf n}_{\mathbf{k}\beta}\cdot\mathbf{e}_{i}^{(j)})e^{-\rm{i}\mathbf{k}\cdot\mathbf{r}_{j}}\vert^{2}
\delta(\omega-\omega_{i}),
\end{equation}
where $\mathbf{e}_{i}^{(j)}$ is the displacement vector of atom $j$ vibrating in eigenmode $i$.  
The transverse dynamical structure factor $S_{T}(\mathbf{k},\omega)$ is not readily 
accessible in scattering experiments, but can easily be calculated from simulation.  
In practice, the orientationally averaged dynamical structure
factor $S_{L}(k,\omega)$ is obtained from scattering experiments \cite{RuoccoSette2001},
and, in this paper, both the orientationally averaged dynamical structure factors 
$S_{L,T}(k,\omega)$ are calculated for $a$-Si.

As observed in experiment, $S_{L}(k,\omega)$ of an amorphous material
exhibits a single peak for low values of wavevector $k$
\cite{RuffleGuimbretiereCourtensVacherMonaco2006,RuoccoSette2001}.  The frequency 
at which this peak is located increases linearly with increasing $k$ 
(at least for low $k$); this is the regime of linear dispersion.  The width of 
this peak also increases with increasing $k$, but more rapidly.  Eventually, the width
of the peak therefore becomes comparable to the peak frequency.  The frequency at which this happens 
(typically $\sim1$ THz) is called the Ioffe-Regel (IR) crossover 
frequency \cite{IoffeRegel1960}.  Above this frequency, 
the mean free path of the wave is shorter than the wavelength, and the external waves do not 
propagate in the conventional sense, instead transferring energy by 
diffusion \cite{AllenFeldmanFabianWooten1999}.

These basic features are undisputed, but large uncertainties in the experimental data
(see, e.g., \cite{CourtensForetHehlenVacher2001}) have meant that it has been difficult
to decide between the two main theoretical models for the behaviour of vibrational plane
waves in amorphous materials.  The first of these \cite{RuoccoSette2001}
suggests that the memory function is the sum of two processes:
structural relaxation, assumed to be too slow to have an effect on vibrational timescales,
and an instantaneous (or at least very fast) process, usually represented as a 
$\delta$-function in time \cite{RuoccoSette2001}.
This leads to $S_{L}(k,\omega)$ having the form of a damped harmonic oscillator (DHO) 
function \cite{RuoccoSette2001}.  Fitting a DHO function to the experimentally observed
peak gives a width $\Gamma$ which scales roughly as the square of the 
wavevector, $\Gamma\propto k^{2}$, for many materials \cite{RuoccoSette2001}.  
This relationship is often seen to hold up to frequencies above the IR crossover.  The
second of these models (see, e.g., \cite{RatForetCourtensVacherArai1999}) is based on an
effective-medium approximation (EMA) \cite{PolatsekEntinWohlman1988} and asserts 
that, in the region of the IR crossover, a DHO function is not a valid fit to
$S_{L}(k,\omega)$, and that the 
scattering becomes much stronger (see, e.g., \cite{RatForetCourtensVacherArai1999}).  
A functional form for $S_{L}(k,\omega)$ is used which imposes 
$\Gamma\propto k^{4}$ by construction, and this has given good fits
of the dynamical structure factor at frequencies close to the IR 
crossover \cite{CourtensForetHehlenVacher2001}.
The behaviour $\Gamma\propto k^{4}$ was also
found recently on fitting a DHO function to $S_{L}(k,\omega)$ of lithium diborate
glass \cite{RuffleGuimbretiereCourtensVacherMonaco2006}.  Recent inelastic 
ultra-violet scattering measurements \cite{Masciovecchioetal2006} on vitreous silica
suggest that there may be three regimes at different frequencies, with $\Gamma\propto k^{2}$, 
$\Gamma\propto k^{4}$ and $\Gamma\propto k^{2}$ respectively.

The simplest possible model for the effect of disorder on vibrations, 
a force-constant-disordered lattice, gives $\Gamma\propto k^{4}$ 
\cite{SchirmacherDiezemannGanter1998,TaraskinLohNatarajanElliott2001}.  A theoretical model
\cite{MartinMayorMezardParisiVerrocchio2001,CilibertiGrigeraMartinMayorParisiVerrocchio2003} 
which includes positional disorder
predicts $\Gamma\propto k^{2}$, but the model is one in which the dynamical
matrix is assumed to be a Euclidean random matrix, and the positions of the atoms
are taken to be uncorrelated and random.  It is not clear how representative
such a model is of real amorphous systems.

Further investigation of the behaviour of the width $\Gamma(k)$ of $S_{L}(k,\omega)$
might therefore be useful to decide on the generality or otherwise of these models.
Investigating $S_{L}(k,\omega)$ by simulation is helpful, because
the large error bars which characterize the experimental results are absent.
Simulated vitreous silica has been shown to have $\Gamma\propto k^{2}$ 
\cite{TaraskinElliott20002,Pillaetal2004} using a DHO function, and fitting
Lorentzian functions to $S_{L}(k,\omega)$ of amorphous silicon ($a$-Si), simulated with the
Stillinger-Weber potential \cite{StillingerWeber1985}, also resulted in a
$\Gamma\propto k^{2}$ dependence for $k\lesssim0.3\textrm{\AA}^{-1}$ \cite{Feldman2002}.
In this paper, we also examine the wavevector dependence of $\Gamma$ in a simulated 
model of $a$-Si, but one simulated using the more physically realistic modified Stillinger-Weber 
potential \cite{VinkBarkemaMousseauvanderWeg2001}, as described in the next section,
as well as fitting to $S_{L}(k,\omega)$
using the DHO fitting function.  Fitting Lorentzian functions to $S_{T}(k,\omega)$
of $a$-Si, produced behaviour suggestive of a $\Gamma\propto k^{2}$ dependence for
$k\lesssim0.3\textrm{\AA}^{-1}$ \cite{FeldmanAllenBickham1999}, but this model was also
constructed with the Stillinger-Weber potential.
The behaviour of both $S_{L}(k,\omega)$ and $S_{T}(k,\omega)$ is investigated here
up to wavevectors of $k\sim2.0\textrm{\AA}^{-1}$, which allows us to show where the DHO
fitting function breaks down, and to investigate the vibrational properties of this
model of $a$-Si at frequencies above the Ioffe-Regel crossover.

\section{Methods}

The model of $a$-Si studied here is a 4096-atom cubic model with box-size $43.274\textrm{\AA}$ 
created by G.~T.~Barkema using a modified WWW method \cite{BarkemaMousseau2000}, and 
then relaxed by us \cite{Christie2006} using the modified Stillinger-Weber (mSW) interatomic 
potential \cite{VinkBarkemaMousseauvanderWeg2001}.  The dynamical
matrix was diagonalized using a Lanczos routine \cite{CullumWilloughby1985}.
The dynamical structure factors were calculated using equation \ref{eq:dsf}.
To remove unimportant physical constants, the modified dynamical structure 
factor $S^{*}_{\beta}(k,\omega)=\frac{m}{k_{B}T}S_{\beta}(k,\omega)$ is plotted 
here, where the polarization $\beta=L$ or $T$,
as in \cite{RuoccoSetteDiLeonardoMonacoSampoliScopignoViliani2000}.
These modified dynamical structure factors were fitted either to a single DHO function with 
two free parameters: peak frequency $\Omega$ and width $\Gamma$, or to a sum of two DHO
functions (for certain wavevectors) with five free parameters: peak frequencies and
widths of the two DHO functions, and the ratio of their intensities.  The peak 
normalization is fixed via $\int\hat{\mathbf{A}}_{\mathbf{k}\beta}(\epsilon)d\epsilon=1$.
To prevent artefacts, all data below 
$\omega=2.5$ meV were ignored in the fitting; the lowest non-zero vibrational 
frequency for this model is $\omega\sim3.7$ meV (at this frequency, the modes
are of mostly transverse character, as seen by the peak in $S_{T}(k,\omega)$
at this frequency in Figure \ref{Tlowk}).  Due to the finite size
of the model, the lowest value of wavevector $k$ compatible with the periodic
boundary conditions is $k=0.145\textrm{\AA}^{-1}$.

Strictly speaking, the dynamical structure factor is a weighted sum of delta-functions, with
non-zero values only at frequencies equal to the mode frequencies (see equation \ref{eq:dsf}).
However, for ease of representation, the delta-functions in the spectral densities were broadened,
before fitting,
by a Lorentzian function with width $\Gamma_{\rm{br}}$.  
The width $\Gamma_{\rm{br}}$ of the broadening chosen has a small, but noticeable, effect
on the value of the width $\Gamma$ of the best-fit DHO function, and hence on the
value of the exponent $\alpha$ in the expression $\Gamma\propto k^{\alpha}$.
$\Gamma_{\rm{br}}$ was usually chosen to be larger than a few energy-level spacings
(the mean energy-level spacing in this model is $\sim0.007$ meV), to smooth
the data effectively for fitting.  However, the larger $\Gamma_{\rm{br}}$ becomes,
the greater is the effect on the raw data.

Increasing the value of $\Gamma_{\rm{br}}$ tends to decrease the value of $\alpha$.
This is to be expected, if we assume that the effect of the Lorentzian broadening
is to increase the true width by an amount comparable to $\Gamma_{\rm{br}}$.
If the true widths are plotted on a double-logarithmic plot, then adding a constant
$\Gamma_{\rm{br}}$ to each width will shift the low-$\Gamma$ points further
up the plot than the high-$\Gamma$ points.  This will have
the overall effect of reducing the gradient of the straight-line fitted to the
broadened data.  A value of $\Gamma_{\rm{br}}$ is required which provides a compromise
between producing broadened data that are smooth enough to be fitted reliably, but 
giving small enough broadening so that the data are not greatly affected.  For the longitudinal
polarization, values of $\Gamma_{\rm{br}}$ in the range $0.03-0.07$ meV were used
(the narrowest peak has width $\sim0.2$ meV), and for the transverse
polarization, values of $\Gamma_{\rm{br}}$ in the range $0.015-0.04$ meV were used
(the narrowest peak has width $\sim0.15$ meV).

\section{Results}
\emph{Longitudinal polarization}

\begin{figure}
\begin{center}
\includegraphics[width=3in,angle=270]{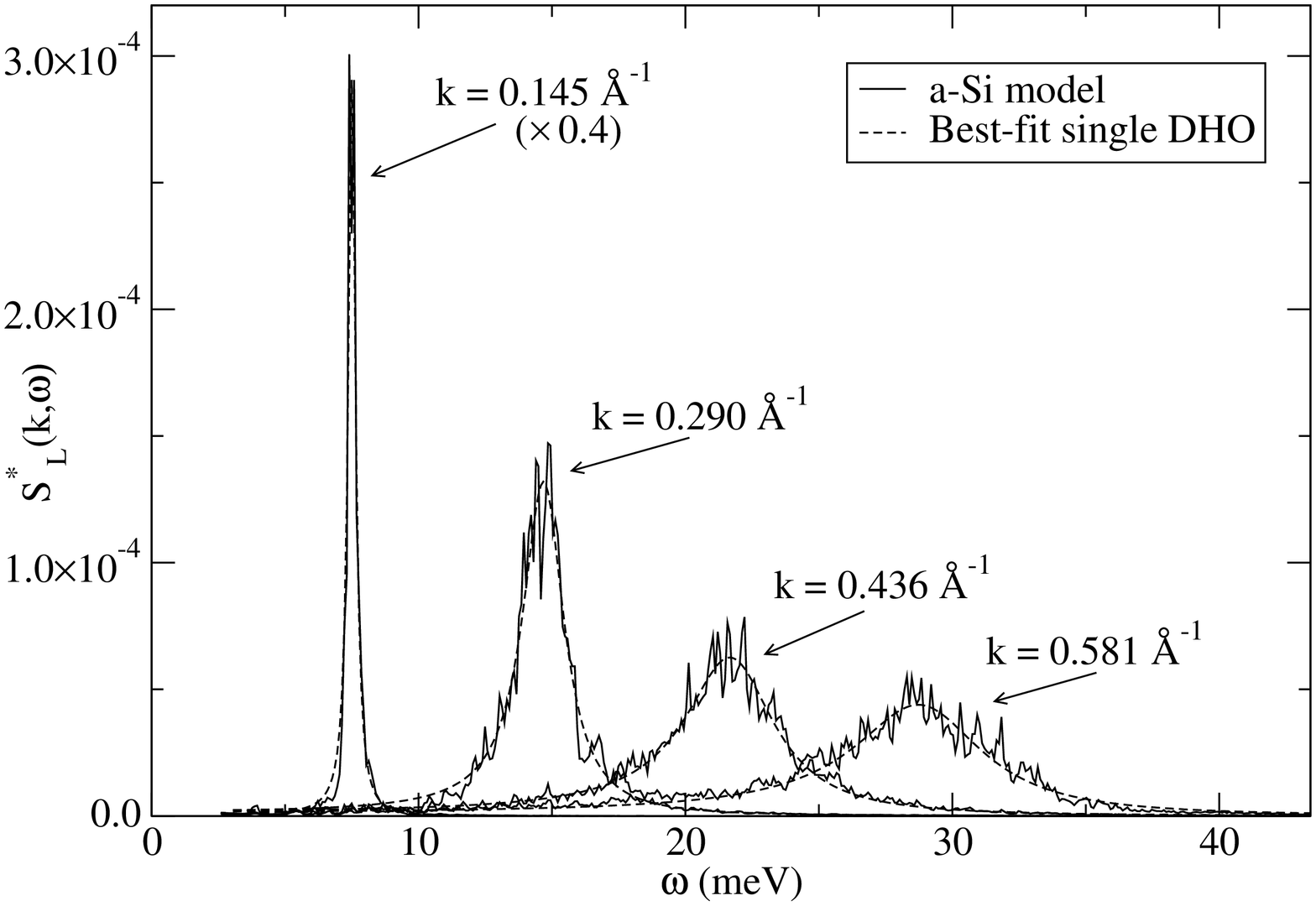}
\end{center}
\caption{\label{L_lowk}
The modified longitudinal dynamical structure factor 
$S^{*}_{L}(k,\omega)$ of the realistic $a$-Si model,
for values of wavevector $k\lesssim0.6\textrm{\AA}^{-1}$, and the 
corresponding best-fit DHO functions. The width $\Gamma_{\rm{br}}$
of the broadening used was 0.05 meV.}
\end{figure}

Figure \ref{L_lowk} shows representative longitudinal modified
dynamical structure factors $S^{*}_{L}(k,\omega)$ for values of 
wavevector $k\lesssim0.6\textrm{\AA}^{-1}$, together with their best-fit DHO functions.  It is clear
that, in this wavevector range, $S^{*}_{L}(k,\omega)$ has only one peak, 
and that, for these wavevectors, the dynamical structure factor is well described
by the DHO function.  The DHO function sometimes deviates slightly from the data,
e.g. on the low-energy side at around $\omega\sim10$ meV of the $k=0.290\textrm{\AA}^{-1}$ peak, 
but these deviations are small.

\begin{figure}
\begin{center}
\includegraphics[width=3in,angle=270]{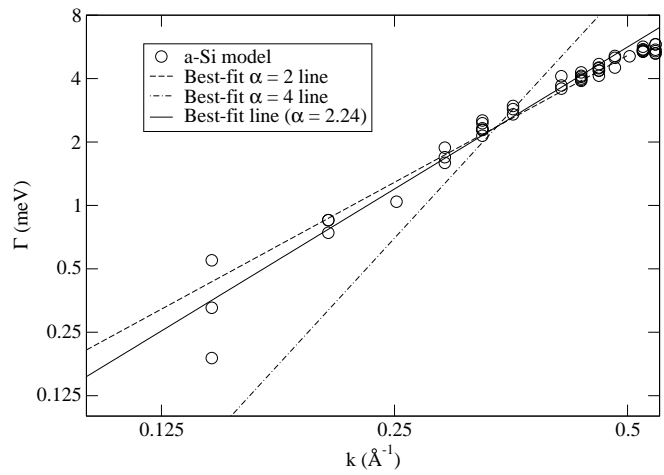}
\end{center}
\caption{\label{Lwidths}
The width $\Gamma$ of the best-fit DHO function to the modified longitudinal dynamical structure
factor $S^{*}_{L}(k,\omega)$
of the realistic $a$-Si model with $\Gamma_{\rm{br}}=0.05$ meV, 
on a double-logarithmic scale, with best-fit $\alpha=2$, $\alpha=4$, and $\alpha$-unconstrained lines, where
$\Gamma\propto k^{\alpha}$, fitted to the data in the range
$0.145<k<0.5\textrm{\AA}^{-1}$.}
\end{figure}

The width $\Gamma$ of the best-fit DHO function is presented in Figure \ref{Lwidths}, as a function of
wavevector $k$, on a double-logarithmic scale, for a representative value of $\Gamma_{\rm{br}}=0.05$ meV.
In Figure \ref{Lwidths}, the data points are seen to lie roughly on a straight line, indicating
that a power-law dependence, $\Gamma\propto k^{\alpha}$, is an appropriate description.
Other values of $\Gamma_{\rm{br}}$ give a width $\Gamma$ which exhibits power-law behaviour
as well.
Figure \ref{Lwidths} also shows the best-fit $\Gamma\propto k^{2}$ $(\alpha=2)$,
$\Gamma\propto k^{4}$ $(\alpha=4)$, and unconstrained $\alpha$ lines to the data in the range
$0.145<k<0.5\textrm{\AA}^{-1}$.  If $\alpha$ is allowed to vary,
then the best-fit value of $\alpha$ for these data is
$2.24\pm0.07$ (this error is estimated by using the sum-of-squares deviation of the data
from the best-fit straight line).  The best-fit $\Gamma\propto k^{2}$ line is shown to be
close to the data,  while $\Gamma\propto k^{4}$ does not represent the data at all.

\begin{figure}
\begin{center}
\includegraphics[%
  width=3in,
  angle=270]{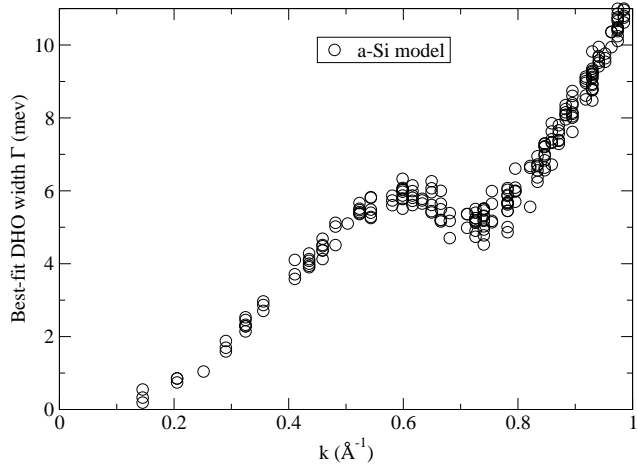}
\end{center}

\caption{\label{widthsto1A-1}The width $\Gamma$ of the best-fit DHO to the modified longitudinal 
dynamical structure factor $S^{*}_{L}(k,\omega)$
of the realistic $a$-Si model with $\Gamma_{\rm{br}}=0.05$ meV.}
\end{figure}

The precise value of $\alpha$ obtained from data such as those in Figure \ref{Lwidths}
depends on the value of broadening $\Gamma_{\rm{br}}$ used, and the range of
wavevectors studied.  It will become clear from the next 
section (Figure \ref{widthsto1A-1}) that the 
power-law behaviour $\Gamma\propto k^{\alpha}$ is not obeyed for 
$k\gtrsim0.55\textrm{\AA}^{-1}$.  Hence data sets were generated 
for values of broadening $0.03<\Gamma_{\rm{br}}<0.07$
meV, and ranges of wavevector starting at 
$0.145<k_{min}<0.26\textrm{\AA}^{-1}$ (to remove possible
finite-size effects), and extending up to 
$0.45<k_{max}<0.55\textrm{\AA}^{-1}$.  For these data sets, 
the mean value of $\alpha$ was 2.08, with a standard deviation of $0.15$.  
98\% of the data sets had $\alpha$ within $\pm0.3$
of this mean.  For no value of broadening $\Gamma_{\rm{br}}$ 
or wavevector range was a line with $\alpha=4$ found to be close to the data.

\begin{figure}
\begin{center}
\includegraphics[%
  width=3in,
  angle=270]{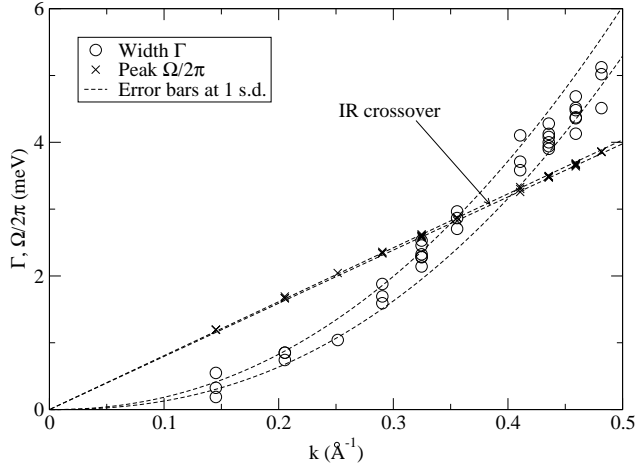}
\end{center}

\caption{\label{IRcrossover_L}The width $\Gamma$ of the best-fit DHO to the 
modified longitudinal dynamical structure factor $S^{*}_{L}(k,\omega)$
of the realistic $a$-Si model, with broadening $\Gamma_{\rm{br}}=0.05$ meV, 
and the scaled
peak frequency $\Omega/2\pi$, plotted against wavevector \emph{k},
with errors to the best-fit lines.  The IR crossover occurs in the region
where the error bars overlap.}
\end{figure}

The Ioffe-Regel (IR) crossover for this model of $a$-Si was estimated, 
defining it to occur at $\Gamma\sim\Omega/2\pi$ \cite{TaraskinElliott2002},
where $\Omega$ is the excitation frequency in $S^{*}_{L}(k,\omega)$.  
For a given value of $\Gamma_{\rm{br}}$ and range of wavevector $k$, an estimate of 
the error in the IR crossover frequency was made, by using the sum-of-squares deviation
from the best-fit straight-line (on a double-logarithmic plot) to obtain 
errors on the value of $\alpha$, and the sound velocity.  As a first approximation,
the IR crossover was assumed to lie in the centre of the overlap region of the 
two sets of error bars.  Figure \ref{IRcrossover_L} displays this procedure 
for $\Gamma_{\rm{br}}=0.05$ meV and wavevectors $0.145<k<0.5\textrm{\AA}^{-1}$.
The longitudinal IR crossover frequency was found to
be $\Omega_{IR,L}=19.1\pm1.3$ meV, corresponding to a wavevector of 
$k_{IR,L}=0.38\pm0.03\textrm{\AA}^{-1}$.  This is somewhat
lower than the previous estimate of $\sim25$ meV \cite{Feldman2002}, due, presumably,
to the different interatomic potential and fitting function used here.
Different values of $\Gamma_{\rm{br}}$ and wavevector range give values of 
$\Omega_{IR,L}$ within the error bars given above.

It follows from this value of $\Omega_{IR,L}$ and Figure \ref{IRcrossover_L} 
that the approximate $\Gamma\propto k^{2}$ behaviour for longitudinal modes for 
this model of $a$-Si continues to frequencies and wavevectors above the IR crossover 
frequency and wavevector, as found in experiment and simulations for other 
materials \cite{RuoccoSette2001}, if the DHO fitting function is used.
There is no sharp change in the behaviour of $S^{*}_{L}(k,\omega)$ evident at or around the IR 
crossover frequency for this model (Figure \ref{L_lowk}), 
nor in the pseudo-dispersion law $\Omega=c_{L}k$, with $c_{L}\sim7670$ m s$^{-1}$ (Figure \ref{IRcrossover_L}).

\begin{figure}
\begin{center}
\includegraphics[width=3in,angle=270]{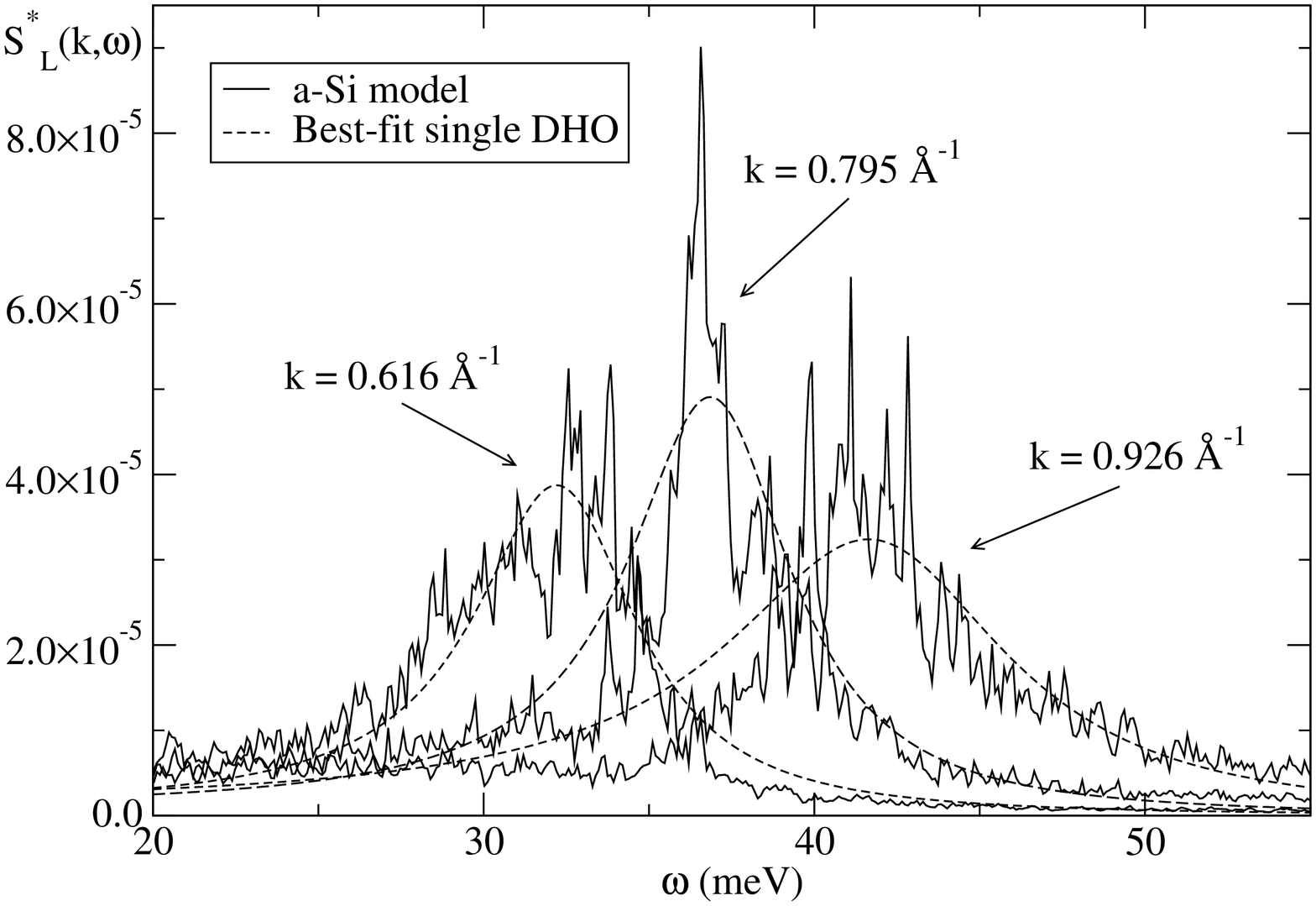}
\end{center}
\caption{\label{L_mediumk}
The modified longitudinal dynamical structure factor 
$S^{*}_{L}(k,\omega)$ of the realistic $a$-Si model,
for values of wavevector $0.6\lesssim k\lesssim1.0\textrm{\AA}^{-1}$, and the best-fit
DHO functions.  The width of the broadening used was $\Gamma_{\rm{br}}=0.05$ meV.}
\end{figure}

\begin{figure}
\begin{center}
\includegraphics[width=3in,angle=270]{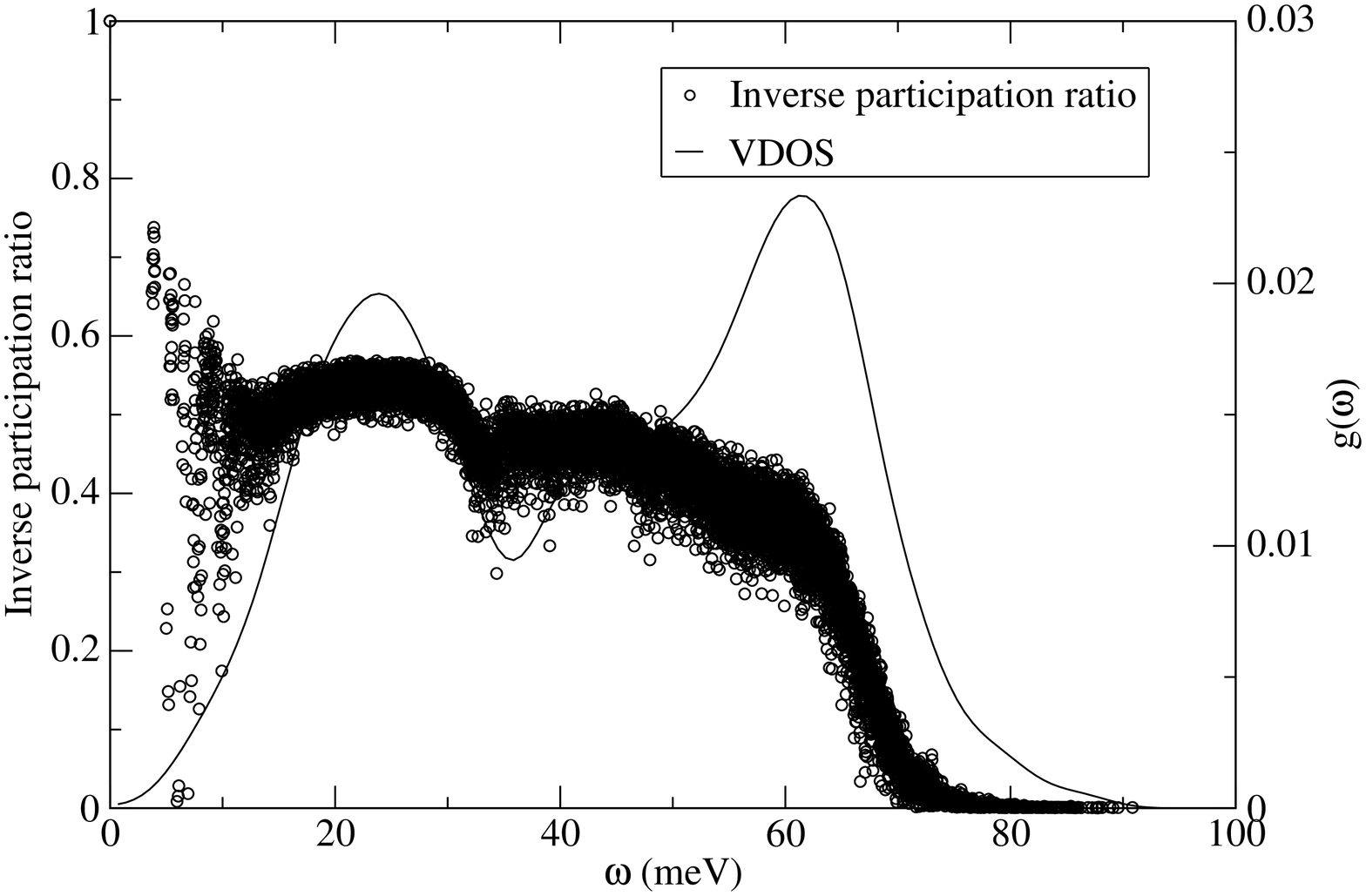}
\end{center}
\caption{\label{IPR}
The inverse participation ratio and vibrational density of states $g(\omega)$
of the realistic $a$-Si model.}
\end{figure}

At higher wavevectors, the DHO fitting function ceases to produce a good fit (Figure \ref{L_mediumk}).  
In addition, for wavevectors in the range $0.55<k<0.8\textrm{\AA}^{-1}$, the peak in 
$S^{*}_{L}(k,\omega)$ narrows (Figure \ref{widthsto1A-1}), although the excitation frequency
continues to increase monotonically (Figure \ref{L_mediumk}).  This narrowing is thought to be related to remnants
of vibrational-mode localization occurring in the gap in the vibrational 
density of states between the acoustic and optic bands, at $\omega\sim35$ meV.  (Peak
excitation frequencies for $0.55<k<0.8\textrm{\AA}^{-1}$ occur in this region.)
This is shown in Figure \ref{IPR}, where the inverse participation ratio of an 
eigenmode, defined as $(\sum_{i}e_{i}^{2})^{2}/N \sum_{i}e_{i}^{4}$, where $e_{i}$ 
is the amplitude of the displacement of atom 
$i$ in that mode, and $N$ is the number of atoms, is shown and is seen to decrease
(indicating greater localization) at roughly the same frequency ($\omega\sim35$ meV)
as the dip in the vibrational density of states between the overlapping bands.
This anomalous dip in $\Gamma(k)$ (Figure \ref{widthsto1A-1}) at $k\sim0.7\textrm{\AA}^{-1}$
is not associated with the IR crossover, which occurs at 
$k_{IR,L}\sim0.38\textrm{\AA}^{-1}$ (Figure \ref{L_mediumk}).

\begin{figure}
\begin{center}
\includegraphics[width=3in,angle=270]{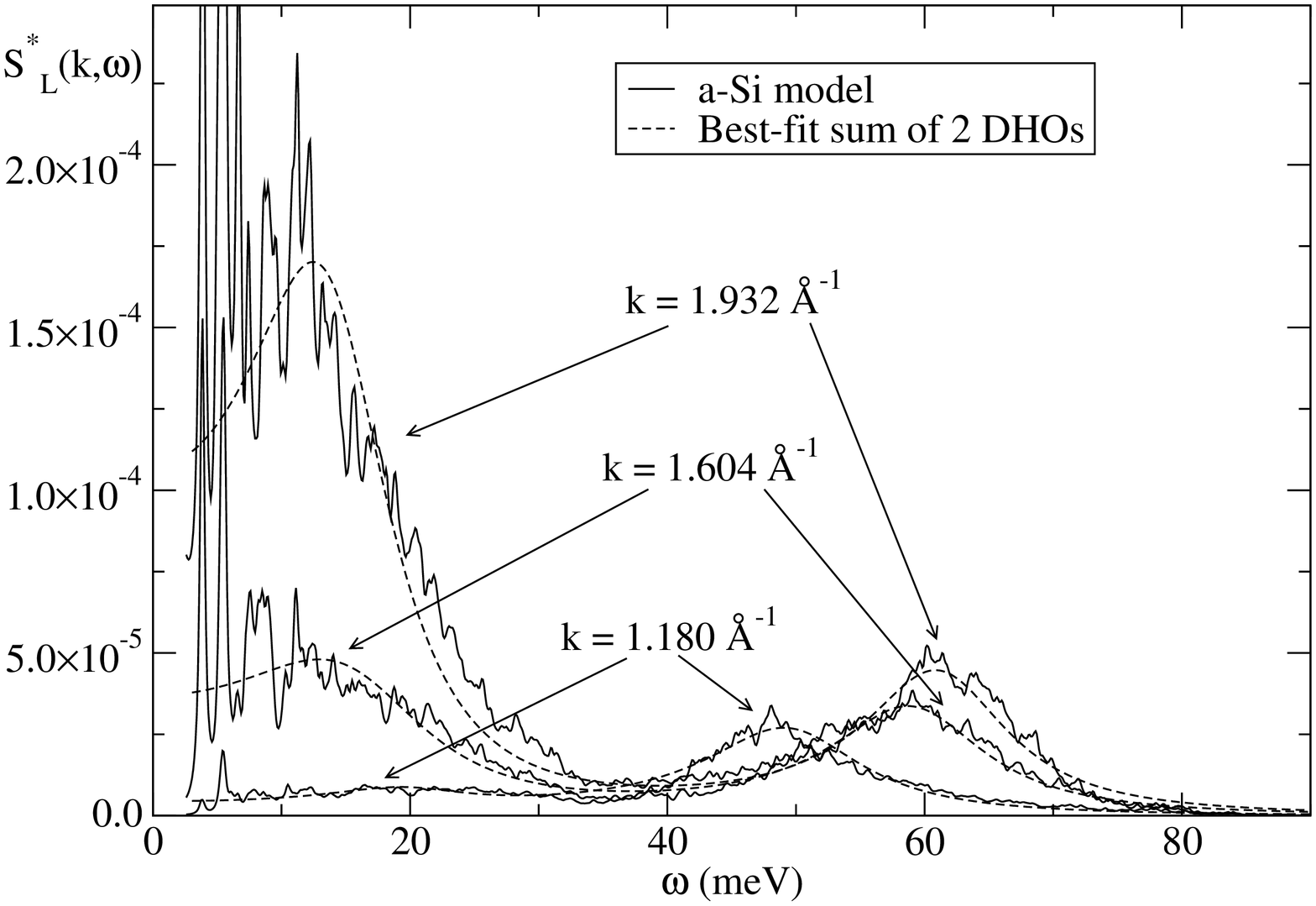}
\end{center}
\caption{\label{Lhighk}
The modified longitudinal dynamical structure factor 
$S^{*}_{L}(k,\omega)$ of the realistic $a$-Si model,
for values of wavevector $k\gtrsim1.0\textrm{\AA}^{-1}$, and the best-fit
sum of two DHO functions.  The value of broadening used was $\Gamma_{\rm{br}}=0.2$ meV.}
\end{figure}

\begin{figure}
\begin{center}
\includegraphics[width=3in,angle=270]{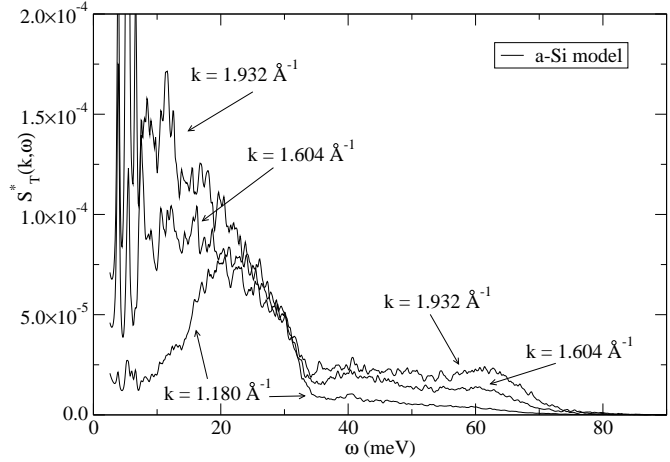}
\end{center}
\caption{\label{Thighk}
The modified transverse dynamical structure factor 
$S^{*}_{T}(k,\omega)$ of the realistic $a$-Si model,
for values of wavevector $k\gtrsim1.0\textrm{\AA}^{-1}$. The value of broadening
used was $\Gamma_{\rm{br}}=0.2$ meV.}
\end{figure}

At even higher wavevectors, $k>1.0\textrm{\AA}^{-1}$, a second peak is evident in
$S^{*}_{L}(k,\omega)$ at low frequency (Figure \ref{Lhighk}).  This peak is not accounted for either in
the memory-function or EMA approaches, but in other materials is thought to be a signature of modes
with a largely transverse character mixing with the longitudinal spectrum, this
disorder-induced mode mixing being a result
of the absence of purely polarized modes in an amorphous material, see, 
e.g., \cite{RuoccoSette2001,TaraskinLohNatarajanElliott2001}.
It will be clear from Figure \ref{Thighk} in the next section that the dominant peak 
in $S^{*}_{T}(k,\omega)$ at high $k$ indeed is responsible for the additional 
low-frequency peak in $S^{*}_{L}(k,\omega)$.  For sufficiently large values of 
wavevector $k$, the spectral density becomes similar to the VDOS \cite{TaraskinElliott2002}.

Figure \ref{Lhighk} also shows the best-fit sum of two DHO functions, which
give reasonable fits to the data. No quantitative analysis of the widths 
was performed for these wavevectors, as the memory-function approach does not 
predict the existence of this second peak, and so fitting the data to two DHO functions 
is largely done \emph{ad hoc}.  Also, for wavevectors $1.0\lesssim k\lesssim1.3\textrm{\AA}^{-1}$
the second peak is small and flat.  The widths found for this peak at these wavevectors 
vary significantly even for data from the same or very similar wavevectors, because 
the second-peak intensity is so small that its width can vary substantially without 
greatly affecting the sum-of-squares deviation.  It
is thus difficult to extract meaningful data about the peak widths for these wavevectors.

\emph{Transverse polarization}

\begin{figure}
\begin{center}
\begin{tabular}{|c|}
\hline
\includegraphics[width=3in,angle=270]{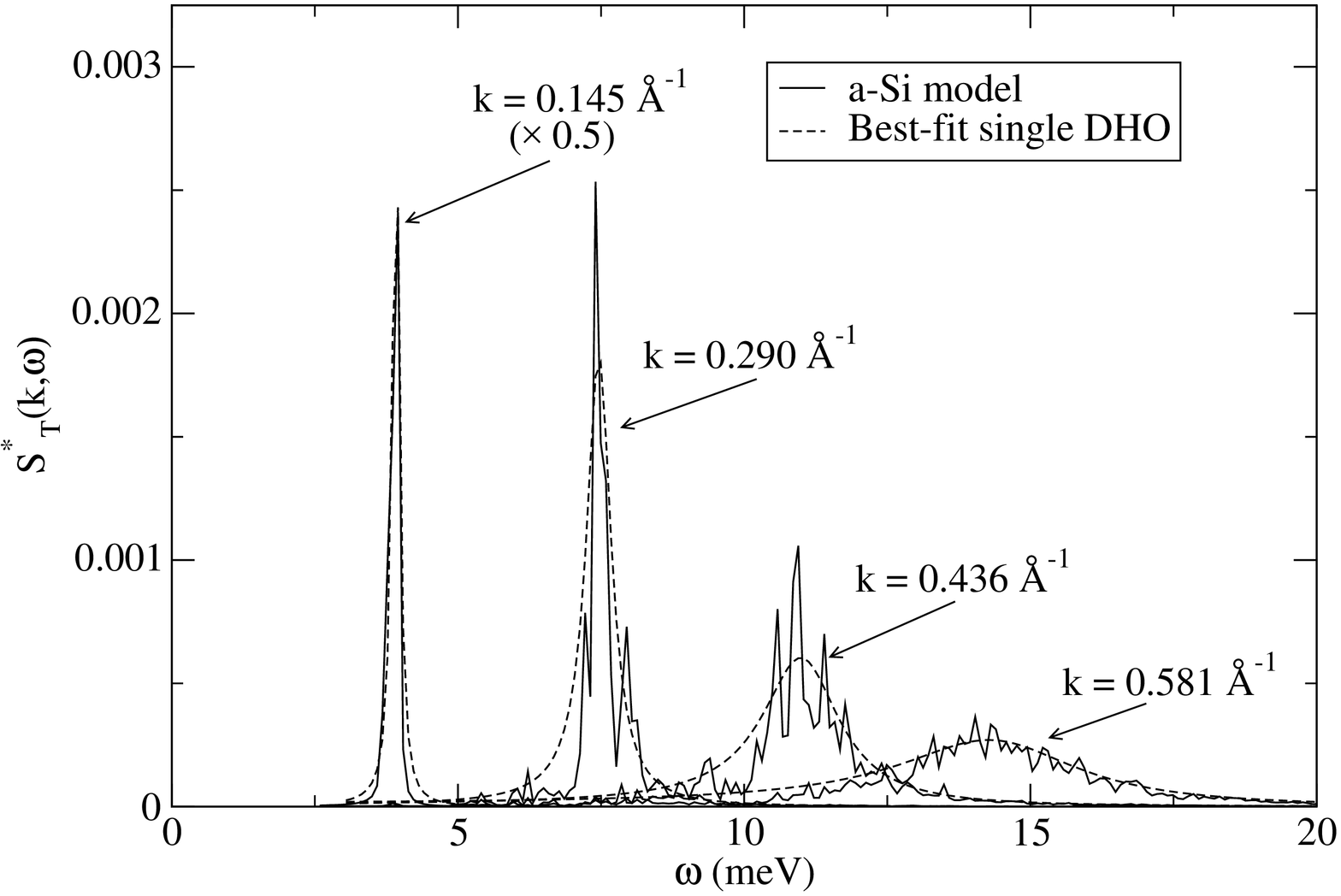}\tabularnewline\hline
\includegraphics[width=3in,angle=270]{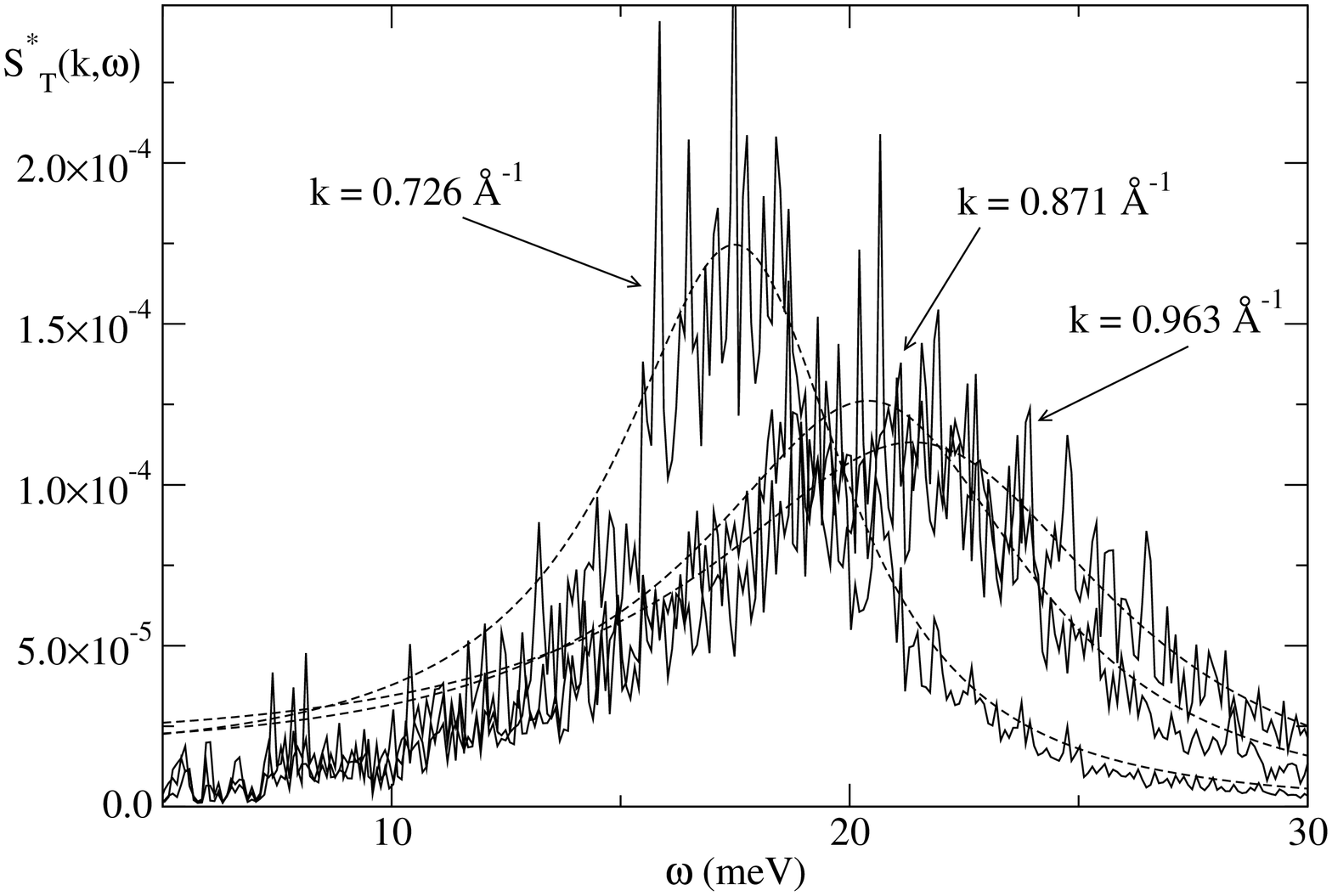}\tabularnewline\hline
\end{tabular}
\end{center}
\caption{\label{Tlowk}
The modified transverse dynamical structure factor 
$S^{*}_{T}(k,\omega)$ of the realistic $a$-Si model,
for values of wavevector $k\lesssim1.0\textrm{\AA}^{-1}$, with the corresponding best-fit
single DHO functions.  The value of broadening used was $\Gamma_{\rm{br}}=0.02$ meV.}
\end{figure}

Figure \ref{Tlowk} shows representative transverse modified
dynamical structure factors $S^{*}_{T}(k,\omega)$, 
and their best-fit DHO functions, for wavevectors 
$k\lesssim1.0\textrm{\AA}^{-1}$.  The width $\Gamma_{\rm{br}}$ of the broadening
used was 0.02 meV.  There is only one peak in $S^{*}_{T}(k,\omega)$ at these frequencies.
The memory-function approach does not predict a particular functional
form for the peak shape in the transverse spectral density, nor in fact does any current 
theory \cite{PontecorvoKrischCunsoloMonacoMermetVerbeniSetteRuocco2005}.
Therefore, different functional forms for the fitting function were tried in an \emph{ad hoc} fashion:
Gaussian, Lorentzian and DHO.  
At very low wavevectors, $k\lesssim0.4\textrm{\AA}^{-1}$, the DHO and Lorentzian functions both gave
good fits to the data.  Above this wavevector, the DHO gave quantitatively the better fit to the data
(in terms of having a lower sum-of-squares deviation), and this remained so for all wavevectors
for which the data could reasonably be described by a single peak 
($k\lesssim1.0\textrm{\AA}^{-1}$).  For this reason, and to facilitate a comparison with
the data for the longitudinal polarization, only the results from the DHO fitting
function are considered here.
From Figure \ref{Tlowk}, it is clear that the DHO function
provides only a reasonable approximation to the data at low wavevectors.
The noise in the data is due in part to the sparseness of eigenvalues at low
frequencies.

\begin{figure}
\begin{center}
\includegraphics[width=3in,angle=270]{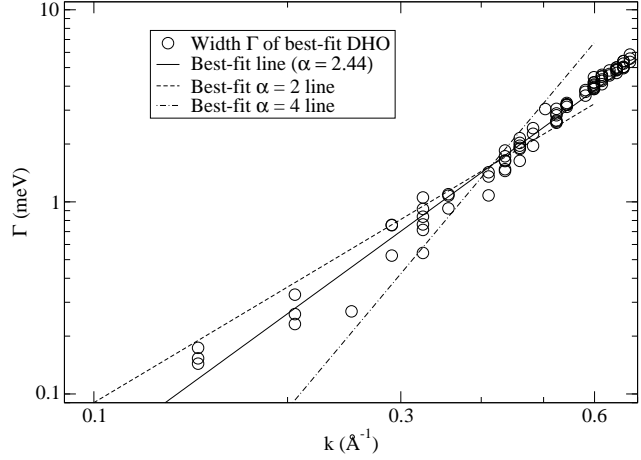}
\end{center}
\caption{\label{Twidths}
The width $\Gamma$ of the best-fit DHO to the modified transverse dynamical structure
factor $S^{*}_{T}(k,\omega)$
of the realistic $a$-Si model, on a double-logarithmic scale, with best-fit 
$\alpha=2$, $\alpha=4$ and unconstrained-$\alpha$
straight lines fitted to the range 
$0.145<k<0.6\textrm{\AA}^{-1}$. The value of broadening used was
$\Gamma_{\rm{br}}=0.02$ meV.}
\end{figure}

As for the longitudinal case, the precise value of the exponent $\alpha$ 
in the dependence $\Gamma\propto k^{\alpha}$ depends on the value
of broadening $\Gamma_{\rm{br}}$ used, and the range of wavevectors studied. 
For the transverse polarization, the physical relevance of the DHO best-fit width $\Gamma$
is less clear, as there is no theoretical reason to expect the DHO function to be
a good fit to $S_{T}^{*}(k,\omega)$.  A discussion of such a fit is included here for completeness.
Figure \ref{Twidths} shows the width $\Gamma$ of the best-fit DHO functions to $S_{T}^{*}(k,\omega)$ as
a function of wavevector $k$, as well as the best-fit lines with $\alpha=2$,
$\alpha=4$ and unconstrained $\alpha$, fitted to the wavevector range 
$0.145<k<0.6\textrm{\AA}^{-1}$, and with $\Gamma_{\rm{br}}=0.02$ meV.  Above 
$k\sim0.6\textrm{\AA}^{-1}$, the DHO function gives a less good fit (Figure \ref{Tlowk}),
and deviations from linear dispersion are also present.  
The value of $\alpha$ in this case is $2.44\pm0.05$.
It is clear that neither $\alpha=2$ or $\alpha=4$ represents the data well, although
$\alpha=2$ is closer.  The precise value of $\alpha$ is likely
to be strongly dependent on the fitting function used.
For values of broadening $0.015<\Gamma_{\rm{br}}<$ 0.04 meV 
and for ranges of wavevector starting at $0.145<k_{min}<0.26\textrm{\AA}^{-1}$, and 
extending to $0.5<k_{max}<0.7\textrm{\AA}^{-1}$, the mean
value of $\alpha$ found was 2.54, with a standard deviation of 0.14.  96\% of
the data sets generated had $\alpha$ within 0.3 of this mean.  For no value of broadening
or wavevector range did the exponent $\alpha=4$ give a good representation of the data.

\begin{figure}
\begin{center}
\includegraphics[%
  width=3in,
  angle=270]{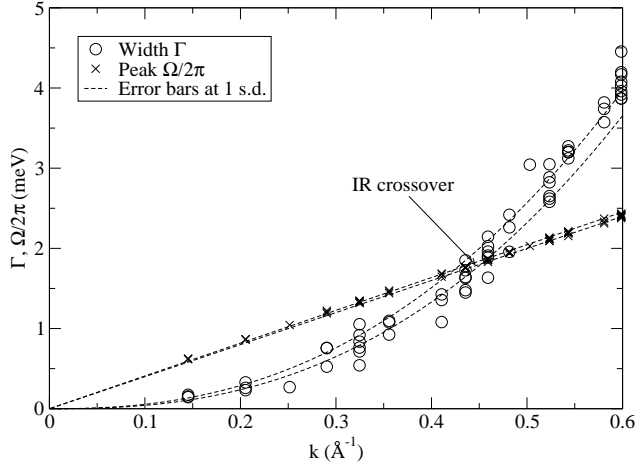}
\end{center}

\caption{\label{IRcrossover_T}The width $\Gamma$ of the best-fit DHO to the 
modified transverse dynamical structure factor $S^{*}_{T}(k,\omega)$
of the realistic $a$-Si model, with broadening $\Gamma_{\rm{br}}=0.02$ meV, 
and the scaled
peak frequency $\Omega/2\pi$, plotted against wavevector \emph{k},
with errors to the best-fit lines.  The IR crossover occurs in the region
where the error bars overlap.}
\end{figure}

Using the best-fit line for $0.145<k<0.6\textrm{\AA}^{-1}$, 
the transverse IR crossover was estimated to be at a
frequency of $\Omega_{IR,T}=11.1\pm0.5$ meV, corresponding to a wavevector of 
$k_{IR,T}=0.44\pm0.02\textrm{\AA}^{-1}$ (Figure \ref{IRcrossover_T}), using the 
same method as for the longitudinal polarization.  This is 
lower than the previous estimate of $\sim17$ 
meV \cite{Feldman2002}, but the different interatomic potential and fitting function 
used in this study may explain this.  As for the longitudinal polarization, 
no sharp change is evident in the width or the shape of $S_{T}^{*}(k,\omega)$ at the
transverse IR crossover (Figure \ref{Tlowk}), nor in the pseudo-dispersion law $\Omega=c_{T}k$, with
$c_{T}\sim3820$ m s$^{-1}$ (Figure \ref{IRcrossover_T}).

For wavevectors above $k\sim1.0\textrm{\AA}^{-1}$, $S^{*}_{T}(k,\omega)$
has one dominant peak at low frequency, with a flat shoulder at higher 
frequencies (Figure \ref{Thighk}).  It is clear that a single peak fit is no longer 
an adequate description of the transverse dynamical structure factor in this wavevector
region.  Neither single DHO nor double DHO fits are very representative of the data 
for these wavevectors.  As mentioned previously, the dominant low-frequency peak in 
$S^{*}_{T}(k,\omega)$ appears in the longitudinal spectrum at high $k$ 
(Figures \ref{Lhighk},\ref{Thighk}), and, vice versa, the high-frequency peak in
$S^{*}_{L}(k,\omega)$ also begins to appear in the transverse spectra, indicative
of strong mode mixing at such high values of $k$.

\section{Discussion}

The width $\Gamma$ of the dynamical structure factor of amorphous materials
increases with increasing wavevector $k$, but there is much dispute about
the functional form of this increase.  Broadly speaking, using a DHO fitting function
(as proposed in \cite{RuoccoSette2001}) to fit to the dynamical structure factor
peaks gives a behaviour of $\Gamma\propto k^{2}$ at low $k$, with the exception of 
lithium diborate, for which $\Gamma\propto k^{4}$ 
\cite{RuffleGuimbretiereCourtensVacherMonaco2006} is found.  A second method,
based on an EMA approach \cite{PolatsekEntinWohlman1988},
which involves modelling the dynamical structure factor with a function which
has $\Gamma\propto k^{4}$ by construction, has also yielded good fits at frequencies close
to the Ioffe-Regel crossover between propagating and diffusive modes.

This paper describes a model of amorphous silicon which is shown to exhibit
an approximate $\Gamma\propto k^{2}$ dependence
for the longitudinal dynamical structure factor.  This relationship
holds for peak frequencies above the IR crossover frequency, as has been found for other
materials modelled with the DHO function \cite{RuoccoSette2001}.  An \emph{ad hoc}
fitting of the transverse dynamical structure factor to a DHO function (without
theoretical justification) gives a slightly higher exponent, about 2.5.  For neither
polarization was any evidence for $\Gamma\propto k^{4}$ found, nor was there
any sharp change in the behaviour of the dynamical structure factor at the IR crossover,
as proposed in the EMA method.

The nature of vibrations in amorphous systems in this frequency region is not yet fully
understood.  It is known that
a force-constant-disordered crystalline lattice gives $\Gamma\propto k^{4}$
\cite{SchirmacherDiezemannGanter1998,TaraskinLohNatarajanElliott2001}, while
many positionally disordered materials, including $a$-Si, give $\Gamma\propto k^{2}$,
as does a positionally disordered analytical model
\cite{MartinMayorMezardParisiVerrocchio2001,CilibertiGrigeraMartinMayorParisiVerrocchio2003}.
It may therefore be positional disorder which is responsible
for the reduction in the exponent from 4 to 2, although the mechanism of this reduction
remains unclear.  Of particular interest is the recent experimental work of Masciovecchio 
\emph{et al.} \cite{Masciovecchioetal2006}, which
suggests that there may be three regimes in vitreous silica, with respectively $\Gamma\propto k^{2}$, 
$\Gamma\propto k^{4}$ and $\Gamma\propto k^{2}$.  If this behaviour is generally true,
then these regions could occur at different ranges of wavevector 
for different materials, and this could account for different dependences being
observed between different materials, e.g. the $\Gamma\propto k^{4}$ dependence in
lithium diborate \cite{RuffleGuimbretiereCourtensVacherMonaco2006}, and for the
intermediate value of exponent $\alpha=2.54$ found for the transverse
polarization above.

\section{Acknowledgements}
We thank G.~T.~Barkema for providing the coordinates of the model. J.~K.~C. is grateful
to the Engineering and Physical Sciences Research Council for the provision of a 
Ph.D. studentship.

\end{document}